# Universal Audio Steganalysis Based on Calibration and Reversed Frequency Resolution of Human Auditory System[1]

Hamzeh Ghasemzadeh[1*], Meisam Khalil Arjmandi[2]

[1] Department of Communicative Sciences and Disorders, Michigan State University, MI, USA
[2] Department of Communicative Sciences and Disorders, Michigan State University, MI, USA
[*]hamzeh_g62@yahoo.com

**Abstract:** Calibration and higher order statistics (HOS) are standard components of image steganalysis. However, these techniques have not yet found adequate attention in audio steganalysis. Specifically, most of current studies are either non-calibrated or only based on noise removal. The goal of this paper is to fill these gaps and to show that calibrated features based on re-embedding technique improves performance of audio steganalysis. Furthermore, we show that least significant bit (LSB) is the most sensitive bit-plane to data hiding algorithms and therefore it can be employed as a universal embedding method. The proposed features also benefit from an efficient model which is tailored to the needs for audio steganalysis and represent the maximum deviation from human auditory system (HAS). Performance of the proposed method is evaluated on a wide range of data hiding algorithms in both targeted and universal paradigms. The results show the effectiveness of the proposed method in detecting the finest traces of data hiding algorithms in very low embedding rates. The system detects steghide at capacity of 0.06 bit per symbol (BPS) with sensitivity of 98.6% (music) and 78.5% (speech). These figures are respectively 7.1% and 27.5% higher than the state-of-the-art results based on RMFCC features.

## 1- Introduction

During the past decade, information security has been revolutionized, and many new trends have emerged. Multimedia encryption systems [1, 2], multimedia secret sharing [3], steganography [4], steganalysis and watermarking [5] are among such trends. Among them, steganography has received a lot of attention. Communicating through a covert channel without arising attention of a third party and preventing traffic analysis are the main purposes of steganography. The outcome of this process is a stego signal ($s \in S$) which results from hiding the intended message ($m \in \mathcal{M}$) inside a host signal, namely called cover ($c \in \mathcal{C}$). Steganography methods can be classified into categories of text, audio, image, video, and network traffics, depending on the type of cover signal.

Steganalysis is the countermeasure of steganography which aims to detect the presence of hidden messages. Likewise, steganalysis methods may be classified according to the type of cover into categories of text, audio, image, video, and network traffics. Steganalysis in each of these categories can be further divided into targeted and universal methods. In the former, the embedding algorithm is known, whereas there is no prior assumption about the embedding algorithm in the later one [6].

One of the first audio steganalysis method was proposed in [7] where cover signal was estimated by de-noising the signal under inspection. Audio quality metrics (AQMs) were used to quantify the discrepancies between the original signal and its estimated cover [7]. Hausdroff distance was proposed as a solution to the inefficiency of AQMs in detecting traces of hidden data [8]. In [9], negative effect of high correlation between the features extracted from these de-noising methods and their signals was solved.

All of these previous works are similar in that, they have used indirect methods for comparing between stegos and their estimated covers. However, conducting this comparison on the distributions of stegos and covers are more appropriate. This approach was pursued in [10], where it was shown that the degree of histograms flatness derived from wavelet coefficients of stegos and their cover counterparts is a discriminative criterion. Gaussian mixture model (GMM) and generalized Gaussian distribution (GGD) were used to capture this criterion. Another improvement was obtained from the second order derivative of audio signal [11]. On the basis of this observation, two different approaches were proposed by incorporating Markov transition probability and Mel-frequency cepstral coefficients (MFCC) [11, 12]. Ghasemzadeh et al. suggested a new steganalysis method by arguing that, by definition, ear should not be able to distinguish between cover and stego signals. According to this argument, MFCC and AQMs are counterintuitive features in steganalysis since their concern is to model ear function. Therefore, to capture hidden messages, an artificial auditory system was proposed which had the maximum deviation from model of ear [13]. Interested reader may refer to [14] for a comprehensive review of audio steganalysis methods.

An important challenge in steganalysis is the tendency of features to show a high degree of within-class variation. This issue makes it hard to find a set of features that are independent from the signal. Different approaches have, therefore, been presented for estimating cover signals to address this issue. These methods that are commonly known as *calibration,* have shown significance improvements in steganalysis performance. Fridrich et al. proposed the first calibration method for JPEG images by recompressing the image after desynchronizing it in the spatial domain [15]. There have been other studies in which high frequency components were noted as the places where information was more likely to be hidden. Ideas of down-sampling [16] and noise-removal [7] for calibration were proposed with respect to this general notion. Down-sampling, as a resolution reduction process preserves the low frequency information while it removes the information within high-frequency regions. Other calibration techniques including recompression [17] and filtering [18] have been also proposed.

There are some shortcomings present in the previous works that we would like to address here. First, although calibration is a common technique in image steganalysis, it has been rarely used for audio steganalysis or it has just been based on noise removal. Second, there is a lack of taking advantage of higher order statistics (HOS) for feature extraction. Third, most of the previous works have not been evaluated thoroughly. That is, mostly they were been examined on LSB methods and mostly in the targeted paradigm. Finally, some of previous works have used cepstrum for capturing characteristics of different frequency bands [11, 19, 20] whereas the actual energy features may be more discriminative. This paper tries to fill these gaps. To this end, we exploit the idea of re-embedding for calibration. This approach is best characterized as a stego estimator instead of a cover estimator. We will extend the applicability of re-embedding to the universal scenario by introducing the notion of bit-plane sensitivity. Using this criterion, it is shown that 1-LSB bit-plane is the most sensitive bit-plane to data hiding algorithms. A universal re-embedding method is proposed on the basis of this observation. After arguing that energy features are more discriminative than their cepstrum counterparts, we propose a new set of features based on a model that we designed previously for steganalysis applications and has the maximum deviation from human auditory system (HAS). Finally, after analyzing different moments, we show that HOS are more discriminative than the first order moment. The proposed system is evaluated on a wide range of data hiding algorithms, including LSB-based, wavelet-based, discrete cosine transform (DCT)-based, and spread spectrum-based and under both targeted and universal paradigms.

The rest of this paper is organized as follows. Section 2 presents the proposed method. Section 3 is devoted to the analysis of the proposed method and extending it to the universal steganalysis paradigm. Simulation results are presented in section 4. Finally, the paper is concluded in section 5.

## 2- The proposed method

Previous studies have shown that taking the second order derivative of audio signal improves the performance of audio steganalysis [11, 12, 19]. Therefore, in the proposed method, the second order derivative of signals is used for feature extraction. After this step, a set of energy-based measures are captured from the segmented chunk of audio signals. The same procedure is applied to the signals embedded with a random message. Steganalysis features, then, are calculated as statistical moments of differences between these measurements. Figure 1 shows the block diagram of the proposed method.

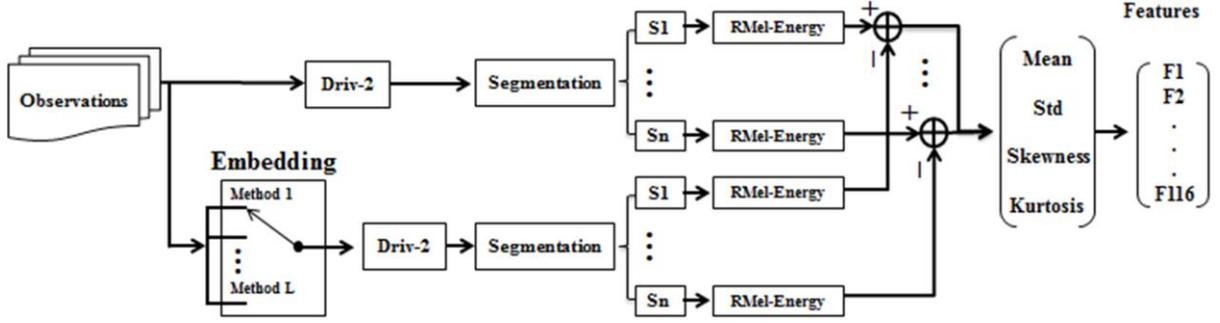

Figure 1. Blockdiagram of the proposed method

### 2.1. Re-embedding calibration

It is possible to use the estimation of stego signal for calibration. Let us take $x$, $\mathcal{A}_{em}$, and $\mathcal{D}$ as the signal under scrutiny, a data hiding algorithm, and a suitable dissimilarity criterion, respectively. Also, we reserve notations of $\mathcal{C}$ and $\mathcal{S}$ for the set of all possible covers and stegos under embedding algorithm of $\mathcal{A}_{em}$. To create the stego signal, the signal $x$ is first embedded with a random message of $r[m]$.

$$\tilde{x}[m] = \mathcal{A}_{em}(x[m], r[m]) \qquad (1)$$

Apparently, $\tilde{x} \in \mathcal{S}$ regardless of $x$. Now, if $x \in \mathcal{S}$:

$$\tilde{x} \in \mathcal{S}, x \in \mathcal{S} \implies \mathcal{D}(x, \tilde{x}) \approx \varepsilon \qquad (2)$$

where $\varepsilon$ is a small value. Also, if $x \in \mathcal{C}$:

$$\tilde{x} \in \mathcal{S}, x \in \mathcal{C} \implies \mathcal{D}(x, \tilde{x}) \not\approx \varepsilon \qquad (3)$$

Using the terminology of pattern recognition, equations 2 and 3 show that re-embedding can reduce intra-class variance and increase inter-class variance; thereby, it can improve the discriminatory strength of the features. According to this observation, the calibrated feature $F$ is calculated as:

$$F = \mathfrak{F}(x[m]) - \mathfrak{F}(\tilde{x}[m]) \qquad (4)$$

where $\mathfrak{F}$ denotes a feature extraction procedure.

### 2.2. Feature extraction procedure

In this section, the proposed feature extraction procedure is presented in detail.

#### 2.2.1 Statistical analysis of energy features

Most of the proposed feature extraction schemes in previous studies relied primarily on some models inspired from HAS. It is a well-identified fact that high frequency regions are more informative for steganalysis [8, 11, 12]. We argue that approaches based on HAS would diminish the performance of steganalysis. To this end, steganography is modeled as an additive and independent noise:

$$s[m] = c[m] + n[m] \qquad (5)$$

where $s$, $c$, and $n$ denote stego, cover, and steganography noise. Dividing the whole spectrum into $L$ equal sub-bands, energy of the signal in each sub-band $B_i$, denoted by $E_i$, is calculated as:

$$(i-1) \times \frac{\pi}{L} \le B_i \le i \times \frac{\pi}{L} \quad, 1 \le i \le L \qquad (6)$$

$$E_i(x[m]) = \int_{B_i} |X(e^{jw})|^2 \, dw \qquad (7)$$

where $X(e^{jw})$ is Fourier transform of $x[m]$. The energy of stego in each sub-band is:

$$E_i(s[m]) = \int_{B_i} |C(e^{jw}) + N(e^{jw})|^2 \, dw \qquad (8)$$

According to [11], when $n[m]$ and $c[m]$ are independent, expected value of (8) is:

$$\int_{B_i} |C(e^{jw})|^2 \, dw + \int_{B_i} |N(e^{jw})|^2 \, dw \qquad (9)$$

Here we define $D_i$ as:

$$D_i = \frac{E_i(s[m])}{E_i(c[m])} = 1 + \frac{\int_{B_i} |N(e^{jw})|^2 \, dw}{\int_{B_i} |C(e^{jw})|^2 \, dw} \qquad (10)$$

$D_i$ can seemingly be used to distinguish between stego and cover signals. Accordingly, $D_i$ value would be one for cover and larger than one for stego signals. Discriminative ability of an energy measure increases as the value of $D_i$ becomes larger. In order to compare the statistical significance of different $D_i$, investigating the power spectral density (PSD) of cover and steganography noise is informative. Figure 2 depicts PSD of a typical speech cover and the noise induced by steghide [21] when it is embedded with a random message.

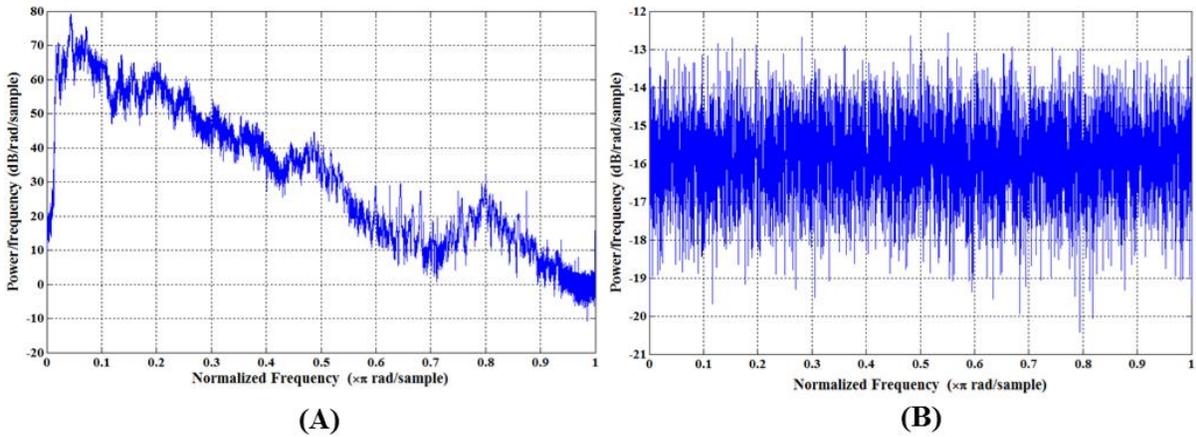

Figure 2. Power Spectral Density (PSD)    (A) A cover signal    (B) Steganography noise, Steghide at 0.12 BPS.

According to figure 2, PSD of the cover signal decreases as its frequency increases, indicating that the cover is a band-limited signal. Comparatively, steganography noise is a wide-band signal which is equally spread over the low and high frequencies. Consequently, if $i>j$:

$$E_i(c[m]) < E_j(c[m]), \qquad (11)$$
$$E_i(n[m]) \approx E_j(c[m])$$

Combining equation (11) with (10) allows us to deduce that energy features are more discriminative as we move toward high frequency regions.

### 2.2.2 Human auditory system and steganalysis

In this section, the conclusion made from the equation (11) is used to show how HAS is counterintuitive for steganalysis applications. Studies on HAS and ears modeling have shown that the sound propagation in the inner ear can be plotted linearly in Mel scale. In these studies, human subjects were asked to listen to a mono tone signal with frequency of $f_1$, and then they were asked to adjust a second signal with the frequency of $f_2$ such that they perceive $f_1/f_2=2$ [22]. Mel scale was the outcome of these subjective experiments. For a given frequency in hertz ($f$), its Mel equivalent is approximated by [23]:

$$Mel = 1127 \times \ln(1 + \frac{f}{700}) \qquad (12)$$

Plotting equation (12) across different frequencies shows that human ear has high frequency resolution in low frequencies and low frequency resolution in high frequencies. Comparing this characteristic with the findings from previous section leads to an interesting conclusion. HAS suppresses high frequency information while it is more sensitive to low frequencies. Therefore, it is not a suitable model for steganalysis purposes. We, consequently, proposed a new scale for audio steganalysis in our seminal work [13]. This new scale was named reversed-Mel and, for a signal with sampling frequency of $F_s$, it was defined as:

$$R - Mel = 1127 \times \ln(1 + \frac{0.5 \times F_s - f}{700}) \qquad (13)$$

Figure 3 compares the frequency resolutions of Mel and R-Mel.

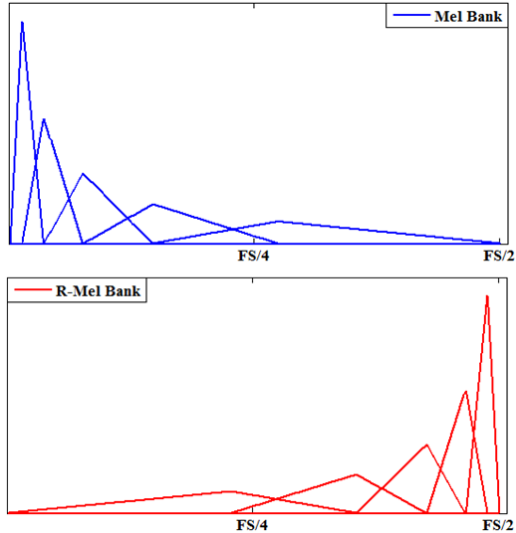

Figure 3.  Triangular filters constructed based on R-Mel and Mel scales

### 2.2.3 Cepstrum and energy features

Cepstrum is anagram of the word spectrum which shows how energy changes in different frequency bands. According to the equation (14), cepstrum is defined as the inverse Fourier transform of the logarithm of power spectrum of the signal [11].

$$cepst_{x[m]} = |F^{-1}\{\log(|F\{x[m]\}|^2))|^2 \quad (14)$$

Similarly, cepstrum can be estimated in another scale such as the proposed R-Mel scale. To this end, after taking Fourier transform of the signal, it is mapped into R-Mel scale using a set of $M$ triangular weighting windows (Figure 3). These windows are constructed as follows:

1. The employed scale is divided into $M+1$ equal sections.
2. The start and stop points of each section is transformed back into hertz scale. Now, we have $M+2$ distinct points.
3. Weighting window $i$ is constructed such that it is zero everywhere except between points $i$ and $i+2$. Also, it is a triangle that raises from point $i$ to pint $i+1$, and then declines from point $i+1$ to point $i+2$.

After that, the logarithm of energy of each filter bank is calculated (equation 15). Then, their inverse Fourier transform are calculated. These final coefficients are called R-MFCCs. Equations (15) and (16) show these steps.

$$E_k = \log\left(\sum F(x[m]).W_k\right), 1 \leq k \leq M \quad (15)$$

$$C_k = |F^{-1}(\log(E_k))| , 1 \leq k \leq M \quad (16)$$

Where $F$, $F^{-1}$, $W_k$, and $M$ are Fourier transform, inverse Fourier transform, triangular weighting windows, and the number of weighting windows, respectively.

The existing audio steganalysis methods are primarily based on cepstrum features (equation 16). This work, instead, uses energy features as defined by equation (15). In the next section, we show that energy features are more discriminative than their cepstrum counterparts.

## 3- Analysis and generalization of the proposed method

### 3.1. Statistical analysis of R-Mel and Mel:

In this section, discriminative ability of R-Mel energy and Mel energy features are compared. Using equations (5) and (15), the difference between the coefficient of cover and stego signal is:

$$\mathcal{D} = \log(\Sigma F(c+n).W_k) \quad (17)$$
$$- \log(\Sigma F(c).W_k)$$

After some basic manipulations, equation (17) reduces to:

$$\mathcal{D} = \log\left(1 + \frac{\Sigma F(n).W_k}{\Sigma F(c).W_k}\right) \quad (18)$$

Logarithm is a monotonic function, that is:

$$x > y \rightarrow \log(x) > \log(y) \quad (19)$$

Therefore, here we may omit the *log* operation for comparison. Apparently, a higher value of $\mathcal{D}$ is more favorable. We want to show that:

$$\frac{\Sigma F(n).W_{k-RMel}}{\Sigma F(c).W_{k-RMel}} \; ? \; \frac{\Sigma F(n).W_{k-Mel}}{\Sigma F(c).W_{k-Mel}} \qquad (20)$$

Assuming the sampling frequency (Fs) of 44.1KHz and *M*=29 and knowing that high frequencies information are more discriminative, we investigate the significance of the coefficient number 29. For this coefficient, the triangular weighting functions of Mel and R-Mel are zero everywhere except in the intervals of [17340, 22050] Hz and [21869, 22050] Hz, respectively. That is, triangle of $W_{29-Mel}$ has a much larger base than that of $W_{29-RMel}$ (figure 3). Therefore, both the numerator and the denominator of right side of equation (19) are much larger:

$$\Sigma F(c).W_{29-Mel} > \Sigma F(c).W_{29-RMel} \qquad (21)$$
$$\Sigma F(n).W_{29-Mel} > \Sigma F(n).W_{29-RMel} \qquad (22)$$

From figure 2.B, it is clear that the energy of steganography noise is much lower than its corresponding cover signal. Therefore, higher value of the numerator (equation 22) cannot compensate for the higher value of the denominator (equation 21). Consequently, the inequality of (19) is correct which implies that R-Mel is more discriminative.

### 3.2. Statistical analysis

To justify our claims that high frequencies and HOS lead to more discriminative features, a set of ANOVA tests was carried out. Signals were segmented, and their spectrums were divided into 50 equal sub-bands. After calculating the energy in each sub-band using equations (6) and (7), their statistical moments were calculated. Figure 4.A presents *p-value* derived from these tests.

To compare the effectiveness of features based on cepstrum coefficients (equation 16) with their energy counterparts (equation 15), 29 coefficients from cover and stego signals were extracted. A set of F-tests, then, was carried out on them. Figure 4.B presents *F-score* from these investigations.

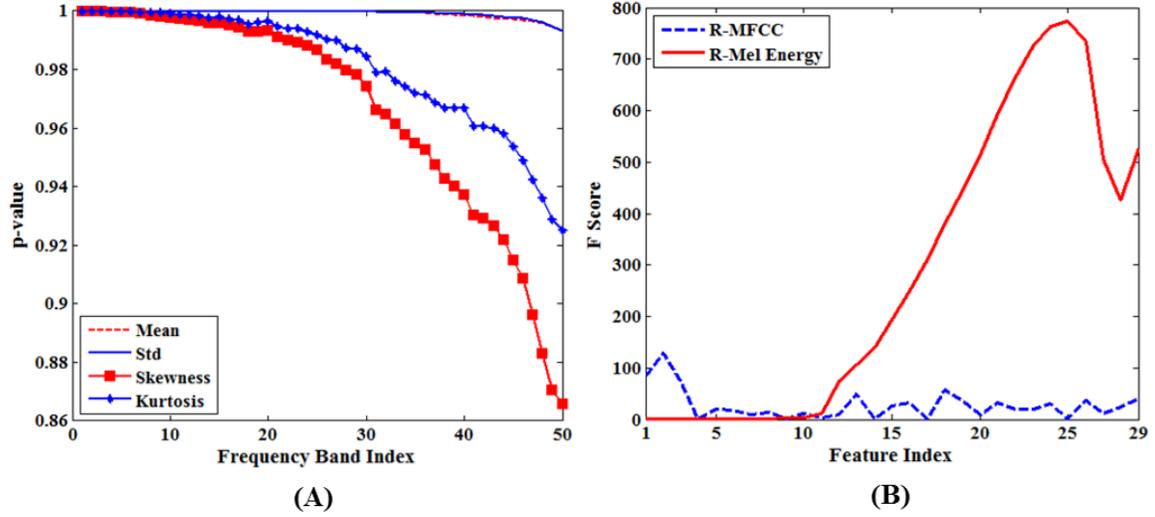

Figure 4. Results of ANOVA test for different feature sets. Stego signals are steghide at 0.12BPS

(A): Significance of HOS and different frequency bands     (B): significance of R-MFCC vs. R-Mel Energy

According to figure 4.A, high frequency bands and HOS result in lower *p-values,* and therefore are more discriminative. Also, according to figure 4.B, features based on R-Mel energy have produced higher scores, and thus are more discriminative.

### 3.3. Generalization

According to figure 1, feature extraction uses re-embedding for calibration. Apparently, the actual embedding algorithm is known in the targeted scenario which allows us to choose it accordingly. However, in the universal scenario, such knowledge is not available which makes us to devise a universal embedding method. Here the noise of steganography is calculated using the following equation.

$$n[m] = s[m] - c[m] \quad (23)$$

We used equation (23) to estimate the probability mass function (PMF) of steganography noise. Our analysis showed that a triangular function can describe this PMF, and also the lower capacities make the PMF sharper. Figure 5 shows PMF of noise of Hide4PGP [24] at different capacities.

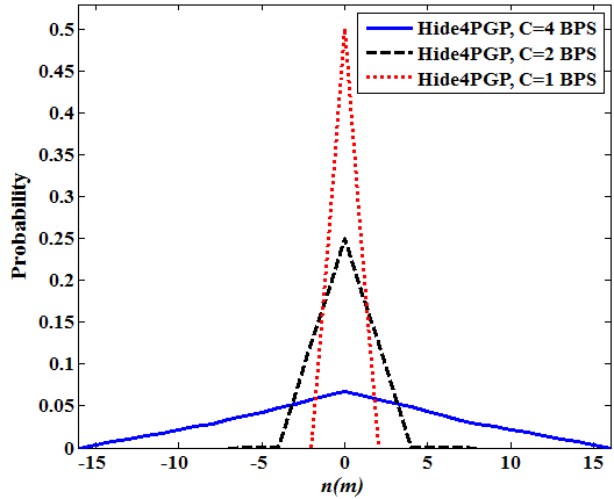

Figure 5. PMF of steganography noise for Hide4PGP at different embedding capacities

According to figure 5, we expect that lower values of noise occur more often than its higher values. Considering that lower values are encoded using LSB and higher values are encoded using other bit-positions, it is expected that the LSB bit-plane be the most sensitive bit-plane to data hiding.

To further justify this, effect of different embedding algorithms on different bit-planes of audio signals were studied. To do this, we define the concept of bit-plane sensitivity. Let $\mathcal{B}_i(x)$ and *BER(x,y)* present, respectively, bit-plane *i* of signal *x* and bit error rate between signals *x* and *y*. Sensitivity of bit-plane *i* of signal *x* is denoted by $\mathbb{S}_i$, which is defined as the amount of noise introduced into bit-plane *i* of cover signal after it is embedded with a random message. Equation (24) shows this.

$$\mathbb{S}_i = 2 \times BER\big(\mathcal{B}_i(c), \mathcal{B}_i(s)\big) \quad (24)$$

For a random message $0 \leq \mathbb{S}_i \leq 1$, where the value of 0 and 1 means that bit-plane *i* is not affected, and it is affected completely with the embedding process, respectively.

A simulation was conducted to measure sensitivities of all bit-planes for different data hiding algorithms. These methods include Hide4PGP [24], Steghide [21], integer to integer wavelet (i2i) [25]. Also, different spread spectrum methods have been proposed in the literature [5, 26-30]. Among them the DCT-based robust watermarking method (COX) [5], spread spectrum watermarking (SSW) [26], spread spectrum in the frequency domain (SS+DCT) [27], and multi carrier spread spectrum (MCSS) as implemented in [30] were tested in this study. Table I presents the average percentage of $\mathbb{S}_i$, $0 \leq i \leq 6$.

TABLE I. BIT-PLANE SESITIVITY(%) OF DIFFERENT EMBEDING METHODS

| Method | Capacity / Param. | LSB POSITION | | | | | |
|---|---|---|---|---|---|---|---|
| | | 1 | 2 | 3 | 4 | 5 | 6 |
| Hide4PGP | C=4 | 100 | 100 | 100 | 95.1 | 0 | 0 |
| | C=2 | 100 | 99.6 | 0 | 0 | 0 | 0 |
| | C=1 | 99.8 | 0 | 0 | 0 | 0 | 0 |
| Steghide | C=0.5 | 49.8 | 24.9 | 12.6 | 6.4 | 3.2 | 1.6 |
| | C=0.25 | 24.9 | 12.5 | 6.4 | 3.3 | 1.6 | 0.8 |
| | C=0.12 | 12.5 | 6.2 | 3.4 | 1.7 | 0.9 | 0.4 |
| | C=0.06 | 6.3 | 3.2 | 1.9 | 1.1 | 0.6 | 0.3 |
| i2i | C=4 | 100 | 100 | 100 | 100 | 100 | 100 |
| | C=2 | 100 | 100 | 100 | 66.4 | 33.3 | 16.8 |
| | C=1 | 100 | 62.4 | 31.2 | 15.7 | 7.9 | 4 |
| | C=0.5 | 50 | 24.9 | 12.4 | 6.2 | 3.1 | 1.5 |
| | C=0.25 | 24.9 | 12.4 | 6.2 | 3.1 | 1.5 | 0.7 |
| | C=0.12 | 12.5 | 6.2 | 3.1 | 1.5 | 0.7 | 0.4 |
| COX | $\alpha = 0.01$ | 100 | 100 | 100 | 100 | 100 | 100 |
| SSW | --- | 99.8 | 99.5 | 99 | 98.2 | 96.8 | 94.4 |
| SS+DCT | a=10 | 99.7 | 99.2 | 97.2 | 90.3 | 71.5 | 43.1 |
| | a=20 | 99.9 | 99.7 | 99.2 | 97.3 | 90.4 | 71.6 |
| MCSS | D=10 | 94.1 | 58.9 | 29.7 | 14.9 | 7.6 | 3.9 |
| | D=20 | 99.9 | 99.8 | 87.7 | 49.6 | 24.9 | 12.6 |

Results in figure 5 and table I show that not only 1-LSB is always affected with data hiding, but it also has the highest sensitivity. On the basis of this observation we propose 1-LSB data hiding as a universal re-embedding method.

## 4- Experiments and results

The proposed feature extraction scheme consists of the following steps:
1- The signal is re-embedded with a random message.
2- Second order derivatives of the signal and its re-embedded version are calculated.
3- Both signals are segmented into frames of 1024 samples with 512 samples overlap.
4- 29 measurements based on R-Mel energy are calculated for every segment of both signals.
5- R-Mel energy of each segment from signal is subtracted from its re-embedded counterpart.
6- Mean ($\mu$), standard deviation ($\sigma$), skewness ($s$), and kurtosis ($k$) of differences are calculated as the steganalysis features.

Let $X$ be a random variable and $E[X]$ denotes its expectation. Equations 25-28 shows the mathematical representation of the employed moments.

$$\mu = E[X] \qquad (25)$$

$$\sigma = \sqrt{E[(X-\mu)^2]} \qquad (26)$$

$$s = E[(\frac{X-\mu}{\sigma})^3] \qquad (27)$$

$$k = \frac{E[(X-\mu)^4]}{(E[(X-\mu)^2])^2} \qquad (28)$$

### 4.1. Experiment Setup:

For the experiments, two different databases were included. The first one is the database used in [13, 19] which contains 4169 wave music clips. The second database was constructed by asking 12 males and 8 females to read a set of Persian articles gathered from daily newspapers in their normal tone and speed. Each session was recorded in our office using a laptop with sampling frequency of 16KHz and resolution of 16 bits. Then, each session was cut into 10 second excerpts. The final database contained 1029 speech wave files.

To generate the stego signals, all cover samples were embedded with all data hiding methods mentioned in section 3.3. Also, to ensure that each stego was embedded with a different message, each message was generated randomly. The same steps were repeated with different message lengths for achieving different capacities.

### 4.2. Performance of the proposed method

To evaluate the performance of different scenarios, we used 10-fold cross validation with support vector machine (SVM). Furthermore, previous works have shown that feature normalization and features selection improve performance of classification [19, 31]. Equation 29 shows the used feature normalization.

$$\widehat{f}_i = \frac{(f_i - m_i)}{\sigma_i} \qquad (29)$$

where $m_i$ and $\sigma_i$ denote values of the mean and the standard deviation of feature $i$ over the train set. Then, the optimum feature set was selected using genetic algorithm (GA). Accuracy was used as the fitness function with a population of 200 and two-point crossover [32]. Table II presents the detailed values of sensitivity and specificity of the proposed method in the targeted scenario.

TABLE II. PERFORMANCE OF THE PROPOSED METHOD IN TERMS OF SENSITIVITY (Se%) AND SPECIFICITY (Sp%)

| Method | Capacity / Param. | Music | | Speech | |
|---|---|---|---|---|---|
| | | Se. | Sp. | Se. | Sp. |
| Hide4PGP | C = 4 | 99.9 | 100 | 100 | 100 |
| | C = 2 | 99.7 | 100 | 100 | 99.9 |
| | C = 1 | 99.7 | 100 | 99.6 | 100 |
| Steghide | C = 0.5 | 99.6 | 100 | 98.2 | 98.2 |
| | C = 0.25 | 99.6 | 99.8 | 90.6 | 91.1 |
| | C = 0.12 | 99.4 | 99.7 | 81.2 | 76.9 |
| | C = 0.06 | 98.6 | 99.6 | 78.5 | 70.2 |
| i2i | C = 2 | 99.8 | 100 | 100 | 100 |
| | C = 1 | 99.8 | 100 | 100 | 99.9 |
| | C = 0.5 | 99.2 | 99.7 | 99 | 98.3 |
| | C = 0.25 | 98.5 | 99.6 | 90 | 90 |
| | C = 0.12 | 99.3 | 99.8 | 78.3 | 73.2 |
| COX | α = 0.01 | 99.6 | 98.5 | 100 | 100 |
| SSW | --- | 99.5 | 99.5 | | |
| SS+ | α = 10 | 100 | 99.9 | 100 | 100 |
| DCT | α = 20 | 99.9 | 99.9 | 100 | 100 |
| MCSS | D = 10 | 99.9 | 100 | 99.2 | 99.8 |
| | D = 20 | 99.9 | 100 | 99.6 | 99.8 |

Performance of the proposed method is compared with some of previous works. Different feature sets were extracted from both databases, and then a set of simulations were carried out.

First, scatter plots of different feature sets are shown as an intuitive measure. Since plots can be drawn at most in 3 dimensions, the most three discriminative features in each set were chosen. Results are presented in figure 6. As shown in this, the distribution of the proposed features has lower intra-class variance and higher inter-class variance.

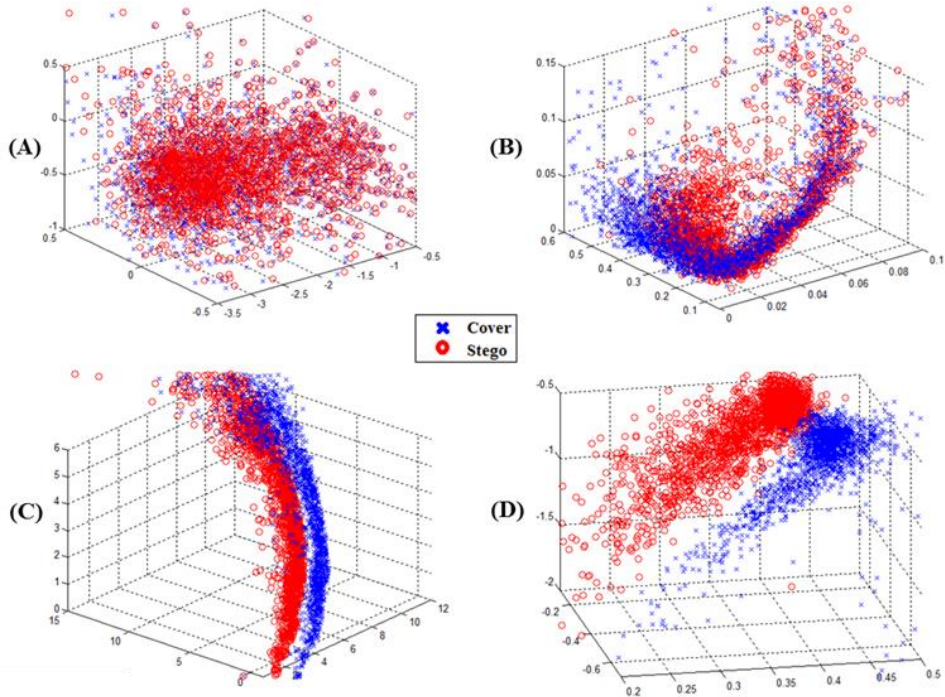

Figure 6. Scatter plot of different feature sets

A: D2-MFCC    B: Markov    C: R-MFCC+GA    D: The Proposed features

To compare performance of different features in the targeted paradigm, they were tested with 10-fold cross validation after normalization. Average performances of the targeted scenario are compared in table III.

TABLE III.    RESULTS OF TARGETED STEGANALYSIS

| Method | Music | | Speech | | Ref |
|---|---|---|---|---|---|
| | Se. | Sp. | Se. | Sp. | |
| **MFCC** | 53.4 | 59.8 | 66.4 | 67 | [20] |
| **D2-MFCC** | 80.2 | 83.2 | 68.8 | 68.3 | [11] |
| **Markov** | 97.4 | 93.3 | 80 | 77.7 | [12] |
| **R-MFCC** | 91.6 | 92.9 | 84.7 | 86.1 | [13] |
| **R-MFCC+GA** | 97.1 | 97.7 | 89.1 | 89.6 | [19] |
| **Proposed** | 99.5 | 99.7 | 94.9 | 93.9 | -- |

As discussed in section 3.2, LSB embedding was adopted in the universal scenario. Results from the universal scenario for the proposed method and some of the previous works are compared in Table IV.

TABLE IV. RESULTS OF UNIVERSAL STEGANALYSIS

| Method | Music | | Speech | | Ref |
|---|---|---|---|---|---|
| | Se. | Sp. | Se. | Sp. | |
| **MFCC** | 42.9 | 78.1 | 48.7 | 96.8 | [20] |
| **D2-MFCC** | 63.9 | 87.6 | 51.4 | 97.8 | [11] |
| **Markov** | 81.4 | 96.1 | 62.9 | 90 | [12] |
| **R-MFCC** | 79 | 94.7 | 66.9 | 98.1 | [13] |
| **R-MFCC+GA** | 87.7 | 98.9 | 71.5 | 97.5 | [19] |
| **Proposed** | 98.7 | 99.8 | 87.8 | 96.7 | -- |

According to tables III and IV, it is evident that the proposed method outperforms previous works by a large margin.

## 5- Conclusion

This study introduced a new set of calibrated features for audio steganalysis based on a model that had maximum deviation from human auditory system. . In the proposed system, the signal was re-embedded with a random message, and then R-Mel energy of original and re-embedded signals were extracted. Eventually, higher order statistics of differences between these measurements were fed to SVM to learn and build the decision boundaries. By investigating PMF of steganography noise and calculating the bit-plane sensitivity of different data hiding algorithms it was argued that 1-LSB embedding can be used as a universal method for calibration. Potency of the proposed system was confirmed on a wide range of data hiding algorithms, including LSB-based, wavelet-based, DCT-based, and spread spectrum based methods under both targeted and universal paradigms.